\documentclass[12pt]{article}
\usepackage{epsfig}

\newcommand{\tlam}{\tilde{\lambda}}
\newcommand{\tR}{\tilde{R}_{\alpha\beta}}
\newcommand{\Dg}{v_{\gamma}}
\newcommand{\ts}{\tilde{\Lambda}}
\newcommand{\s}{\Lambda}
\newcommand{\tT}{\tilde{\cal T}}
\newcommand{\tU}{\tilde{U}}
\newcommand{\tJ}{\tilde{J}}
\newcommand{\tP}{\tilde{P}}
\newcommand{\tcA}{\tilde{\cal A}}
\newcommand{\tA}{\tilde{A}}
\newcommand{\tB}{\tilde{B}}
\newcommand{\Kijab}{K^{ij}_{\alpha\beta}}
\newcommand{\tK}{\tilde{K}^{ij}_{\alpha\beta}}
\newcommand{\tDel}{\tilde{\Delta}}
\newcommand{\tX}{\tilde{X}}
\newcommand{\tH}{\tilde{H}}

\newcommand{\tplaq}{\tilde{U}_{\alpha i}\tilde{U}^*_{\beta i}\tilde{U}^*_{\alpha j}\tilde{U}_{\beta j}}
\newcommand{\tp}{\tilde{H}}
\newcommand{\tq}{\tilde{Q}}
\newcommand{\HHH}{H_{e\mu}H_{\mu\tau}H_{\tau e}}
\newcommand{\tHHH}{\tH_{e\mu}\tH_{\mu\tau}\tH_{\tau e}}

\newcommand{\discr}{\Delta_{12}\Delta_{23}\Delta_{31}}
\newcommand{\tdiscr}{\tDel_{12}\tDel_{23}\tDel_{31}}

\newcommand{\beq}{\begin{equation}}
\newcommand{\eeq}{\end{equation}}
\newcommand{\bea}{\begin{eqnarray}}
\newcommand{\eea}{\end{eqnarray}}

\newcommand{\nubar}{{\bar \nu}}
\newcommand{\numu}{\nu_\mu}
\newcommand{\nutau}{\nu_\tau}
\def\ni{\noindent}
\def\nl{\hfill\break}
\begin{document}
\begin{flushright}
hep-ph/0305175 \\
RAL-TR-2003-014 \\
15 May 2003 \\
\end{flushright}
\begin{center}
{\Large Exact Matter-Covariant Formulation of}
\end{center}
\vspace{-7mm}
\begin{center}
{\Large Neutrino Oscillation Probabilities}
\end{center}
\vspace{1mm}
\begin{center}
{P.F.~Harrison\\
Physics Department, Queen Mary University of London\\
Mile End Rd. London E1 4NS. UK \footnotemark[1]}
\end{center}
\begin{center}
{and}
\end{center}
\begin{center}
{W.G.~Scott\\
CCLRC Rutherford Appleton Laboratory\\
Chilton, Didcot, Oxon OX11 0QX. UK \footnotemark[2]}
\end{center}
\begin{center}
{and}
\end{center}
\begin{center}
{T.J.~Weiler\\
Department of Physics \& Astronomy, Vanderbilt University\\
Nashville, TN 37235 USA \footnotemark[3]}
\end{center}
\vspace{1mm}
\begin{abstract}
\baselineskip 0.6cm
\noindent

We write the probabilities for neutrino oscillations in uniform-density matter 
exactly in terms of convention-independent vacuum neutrino
oscillation parameters and the matter density.
This extends earlier results formulating neutrino oscillations
in terms of matter-, phase-, and trace-invariant quantities.
\end{abstract}

\vspace{2cm}
\begin{center}
{\it To be published in Physics Letters B}
\end{center}

\footnotetext[1]{E-mail:p.f.harrison@qmul.ac.uk}
\footnotetext[2]{E-mail:w.g.scott@rl.ac.uk}
\footnotetext[3]{E-mail:tom.weiler@vanderbilt.edu.us}

\newpage
\ni {\bf 1 Introduction}
\vspace{2mm}
\nl 
The advent of oscillation experiments in which neutrinos traverse 
significant distances in terrestrial matter before observation 
make the relationship between 
the vacuum oscillation parameters and the oscillation observables in 
matter of considerable interest. Some of us recently pointed out 
\cite{matterInv1}\cite{matterInv2} the significance here 
of matter-invariants: neutrino oscillation parameters which are invariant 
under the influence of matter may be used to simplify the relationship 
between neutrino oscillation observables and the vacuum neutrino 
oscillation parameters.

The effective neutrino oscillation Hamiltonian in the flavour
basis (the weak basis in which the charged lepton mass matrix is
diagonal) is given by $H=MM^{\dag}/2E$, where $M$ is the neutrino mass 
matrix and $E$ is the neutrino energy. In vacuum, this is diagonalised by
the conventional MNS mixing matrix \cite{MNS}, $U$: 
$U^{\dag}HU= {\rm diag}(\lambda_1,\lambda_2,\lambda_3)$, where the 
$\lambda_i=m_i^2/2E$ are the vacuum eigenvalues. 
The effects of matter on the propagation of
neutrinos is described by the addition of the 
Wolfenstein term, $a=\sqrt{2}G_F N_e$ ($N_e$ is the number density of 
electrons in the matter), to the $H_{ee}$ element:
\begin{equation}
\tilde{H}=H+{\rm diag}(a,0,0)
\label{matterHam}
\end{equation}
which modifies the eigenvalues and the elements of the MNS matrix in
a non-trivial way.  We denote 
matter-modified parameters by quantities with a $\tilde{~}$.

The electron density in matter is not $C$-, $CP$- or $CPT$-symmetric:
for antineutrino propagation, $a$ in Eq.~(\ref{matterHam}) keeps its magnitude 
but changes sign. Thus, the matter-modified eigenvalues and mixing matrix 
for antineutrinos are different from those for neutrinos. 
We treat the general case, leaving the matter density, $a$, a free parameter, 
and comment further on the relationship between the neutrino and antineutrino 
cases in Appendix A.

The formula for the 
appearance and survival probabilities as a function of propagation distance, 
$L$, when neutrinos pass through uniform density matter may be written 
in its usual form, but in terms of the matter-modified parameters 
as follows:
\begin{eqnarray}
\tP(\nu_{\alpha} \rightarrow \nu_{\beta})
&=& \delta_{\alpha\beta}
-4\sum_{i<j}\tK\sin^2{(\tDel_{ij}L/2)} \cr
&+&8\tJ_{\alpha\beta}\sin{(\tDel_{12}L/2)}\sin{(\tDel_{23}L/2)}\sin{(\tDel_{31}L/2)}
\label{prob}
\end{eqnarray}
where the 
\begin{equation}
\tK={\rm Re}(\tU_{\alpha i}\tU^*_{\beta i}\tU^*_{\alpha j}\tU_{\beta j}),
\label{Kdef}
\end{equation}
parameterise the magnitudes of the $T$-even oscillations and
\begin{equation}
\tJ_{\alpha\beta}
={\rm Im}(\tplaq)
=\pm \tJ~({\rm for}~\alpha \neq \beta)~{\rm or}~=0~({\rm for}~\alpha=\beta)
\label{Jdef}
\end{equation}
parameterises the magnitude of the $T$-odd oscillations\footnote
{We prefer the ``$T$-even'' and ``T-odd'' labels to the ``$CP$-even'' and 
``$CP$-odd'' ones, since matter introduces an extrinsic $CP$-odd contribution into the intrinsically CP-even terms, while spherically-symmetric matter profiles respect $T$-invariance for neutrino propagation between points at
equal radii (eg.~on the surface of the Earth).}.
The eigenvalue differences in matter, 
$\tDel_{ij}\equiv (\tilde{m}^2_i -\tilde{m}^2_j)/2E$ may be calculated in terms of the vacuum parameters, $\Delta_{ij}$ and $U_{\alpha i}$, and the Wolfenstein term, $a$, using the solutions of the cubic characteristic equation of the matter-modified Hamiltonian \cite{bargeretal}\cite{zaglauer}. The matter-modified MNS matrix elements, $\tU_{\alpha i}$, may be similarly calculated, but are rather complicated functions \cite{zaglauer} of the vacuum parameters and the Wolfenstein term. It is the aim of this paper to simplify as much as possible the relationship between the observable oscillation amplitudes in matter, $4\tK$ and $8\tJ$, and the vacuum parameters.

\vspace{7mm}
\ni {\bf 2 Matter Invariance}
\vspace{2mm}
\nl 
The idea of matter-invariance is based on the observation that 
all quantities $\tH_{\alpha\beta}$ in Eq.~(\ref{matterHam}) other than 
$\tH_{ee}$ are 
matter-invariant, and appropriately combined, can be related to 
observable parameters ($H_{ee}$, because of its trivial transformation 
in matter, may be said to be ``matter-covariant''). 
The first application of these ideas showed that Jarlskog's 
determinant \cite{jarlskog} was matter-invariant and led to the so-called NHS relation \cite{matterInv1}:
\begin{equation}
\tdiscr\tJ={\rm Im}(\tHHH)
={\rm Im}(\HHH)=\Delta_{12}\Delta_{23}\Delta_{31}J.
\label{ImHHH}
\end{equation}
This relation was used to write exactly the $T$-violating part of the matter-modified oscillation probability, Eq.~(\ref{prob}), in a very simple and compact form:
\begin{equation}
\tilde{P}_{\not T}(\nu_{\alpha} \rightarrow \nu_{\beta})
=8\frac{\discr}{\tdiscr}J
\sin{(\tilde{\Delta}_{12}L/2)}\sin{(\tilde{\Delta}_{23}L/2)}
\sin{(\tilde{\Delta}_{31}L/2)}
\quad (\alpha\neq\beta).
\label{triple3}
\end{equation}
The advantage of this formulation is that it does not require the matter-modified MNS matrix elements, whose expressions in terms of vacuum parameters are quite complicated \cite{zaglauer}. The matter-dependence is
confined to the eigenvalue differences, $\tDel_{ij}$, for which it is somewhat more straightforward \cite{bargeretal}\cite{zaglauer}.

The matter-invariant approach was subsequently extended \cite{matterInv2} to the $T$-even part of the oscillation probability, for which analogous (but less
simple) expressions to Eq.~(\ref{ImHHH}) were found. Analogues of 
Eq.~(\ref{triple3}), valid in approximation were obtained, but exact 
formulations were not.

The matter-invariant approach was next applied ingeniously \cite{kimura} 
to provide, after a lengthy derivation, exact expressions for the 
$T$-conserving coefficients \cite{ohlsson} in terms of the effective 
Hamiltonian elements and the matter-dependent eigenvalues: 
\begin{equation}
\tK
=\frac{|\tp_{\alpha\beta}|^2\tlam_i\tlam_j+|\tq_{\alpha\beta}|^2
+{\rm Re}(\tp_{\alpha\beta}\tq^*_{\alpha\beta})(\tlam_i+\tlam_j)}
{\tDel_{12}\tDel_{23}\tDel_{31}\tDel_{ij}}
\quad (\alpha \neq \beta)\,.
\label{kimura}
\end{equation}
Here $\tq^*_{\alpha\beta}$ is the cofactor of $\tH_{\alpha\beta}$ and is matter-invariant for $\alpha$ or $\beta=e$, while $\tp_{\alpha\beta}$ is matter-invariant for $\alpha$ or $\beta\neq e$. For the two off-diagonal cases with $\{\alpha,\beta\}=\{e,\mu\}$ or $\{e,\tau\}$ (the $\tK$ are symmetric
under the interchange of $\alpha$ and $\beta$), this approach isolates the
matter-dependence to the $\tlam_i$ alone, in the same spirit as, eg.~Eq.~(\ref{triple3}) isolates it to the $\tDel_{ij}$. However, for the 
case $\{\alpha,\beta\}=\{\mu,\tau\}$, $\tq_{\alpha\beta}$ is not 
matter-invariant, which complicates the formulation, and spoils the 
symmetry of the formulation between the different components.

Neutrino oscillation observables can depend only on the differences between mass-squared eigenvalues and so must be ``trace-invariant,'' ie.~invariant under transformations in which any multiple of the identity is added to the effective neutrino Hamiltonian; a change in the trace is equivalent to a change in the overall phase of the neutrino propagation amplitude. Observables are, in addition, ``phase-invariant'', ie.~invariant under phase transformations of the neutrino mass eigenstates. Hence, it must be possible to write the relationship between the observables and the vacuum parameters entirely in terms of trace- and phase-invariant quantities. Eqs.~(\ref{ImHHH}) and (\ref{triple3}) are examples of this. While the particular combination given in Eq.~(\ref{kimura}) is of course both trace- and phase-invariant, this formulation suffers the difficulty that neither the $\tq_{\alpha\beta}$ nor the $\tlam_i$ are trace-invariant. These individual quantities cannot therefore be related to observables of neutrino oscillations. Before the values of the $\tq_{\alpha\beta}$ or the $\tlam_{i}$ can be specified, an artificial offset of the neutrino masses must be chosen. In the applications cited in \cite{kimura} the offset is arbitrarily set so that $m^2_1=0$ in vacuum.

In the remainder of this paper, we provide a unified formulation, using 
matter-invariants which are trace- and phase-invariant. The 
matter-dependence is isolated in factors which depend only on the 
eigenvalue differences, $\tDel_{ij}$, and the matter density itself. We 
find the exact $T$-even analogues of Eqs.~(\ref{ImHHH}) and 
(\ref{triple3}), and hence exact convention-independent, matter-covariant 
expressions for the observable neutrino oscillation probabilities in 
terms of vacuum parameters.

\vspace{7mm}
\ni {\bf 3 Matter-Covariant Derivation of Oscillation Probabilities in 
Uniform Density Matter}
\vspace{2mm}
\nl 
We provide a matter-covariant derivation of the neutrino oscillation
probabilities given in Eq.~(\ref{prob}). The amplitude $\tcA_{\alpha\beta}$
for a neutrino of flavour $\alpha$ to be detected as a neutrino of flavour 
$\beta$ in matter of uniform density is given as a function of propagation distance $L$ by the (matrix) equation:
\begin{equation}
\tcA = \exp(-i\tH L)
\label{expHam}
\end{equation}
where $\tH$ is the effective neutrino oscillation Hamiltonian of Eq.~(\ref{matterHam}). The general theory for a function of an operator \cite{shaw} enables the exponentiation to be performed directly in the flavour basis:
\begin{equation}
\tcA = \sum_i \tX^i \exp(-i\tlam_i L)
\label{amplitude}
\end{equation}
where the $\tlam_i$ are the eigenvalues of $\tH$
and the Hermitian projection operators $\tX^i$ are given by:
\begin{eqnarray}
\tX^i 
&=&\frac{\prod_{j \neq i} (\tH-\tlam_j)}
                {\prod_{j \neq i} (\tlam_i-\tlam_j)}\label{lagrange}\\
&=&\frac{(\tH-\tlam_j)(\tH-\tlam_k)}
{\tDel_{ij}\tDel_{ik}} \quad (j \neq k \neq i)\,.
\label{lagrange2}
\end{eqnarray}
The second equality is specific to three families of neutrinos. Comparison with the more conventional approach in which the Hamiltonian is diagonalised before exponentiation allows us to identify the elements of the $\tX^i$ with the familiar combinations of the lepton mixing matrix elements \cite{HPS33}:
\begin{equation}
\tX^i_{\alpha\beta} = \tU_{\alpha i}\tU^*_{\beta i}
\label{xi}
\end{equation}
(no summation over $i$ is implied). The $\tX^i_{\alpha\beta}$
for $\alpha\neq\beta$ are not phase-invariant, and are therefore
not observables (although the diagonal elements
$\tX^i_{\alpha\alpha}=|\tU_{\alpha i}|^2$ are). All components of the 
$\tX^i$ are however trace-invariant, this property being manifest 
in the combinations $(\tH-\tlam_i)$ which appear in Eqs.~(\ref{lagrange}) 
and (\ref{lagrange2}) (or in the fact that diagonalization of $\tH$,
yielding the $\tU$'s appearing in Eq.\ (\ref{xi}),
is a trace-invariant process).

The neutrino oscillation probabilities of Eq.~(\ref{prob}) are
given by the squared amplitude
$\tP(\nu_{\alpha} \rightarrow \nu_{\beta})=|\tcA_{\alpha\beta}|^2$,
which contains real and imaginary projections, 
$\tK$ and $\tJ$ respectively, of the  
phase-, trace-, and convention-independent plaquettes \cite{bjorken}:
%
\begin{equation}
\tplaq=
\tX^i_{\alpha\beta}\tX^{j*}_{\alpha\beta} \quad(i\neq j) \,.
\label{plaq}
\end{equation}
Comparing Eqs.~(\ref{xi}) and (\ref{plaq}), one sees that the 
$\tX^i_{\alpha\beta}$ represent an intermediate calculational step 
between the $\tU_{\alpha i}$ and the observable plaquettes.  
Unlike the $\tU_{\alpha i}$ and the plaquettes however,
the $\tX^i_{\alpha\beta}$ have a simple unitarity relation: 
$\sum_i \tX^i_{\alpha\beta}=\delta_{\alpha\beta}$.
The $\tX$'s close the relation among the fundamental parameters 
$\tH$ and $\tlam$: the relations
\beq
\tlam_i = Tr(\tX^i \tH)\,,
\label{set1}
\eeq
and 
\beq
(\tH-\tlam_k)=\sum_i \tDel_{ik}\tX^i\quad {\rm for~given}~k,
\label{set2}
\eeq
together with Eq.~(\ref{lagrange2}) 
demonstrate the equal status of the 
$\tX$, $\tH$, and $\tlam$. Any pair of these sets of quantities encapsulates
equivalent information. 

We now proceed to develop Eq.~(\ref{lagrange2}) by explicitly 
calculating the $\tX^i_{\alpha\beta}$ in terms of the $\tH$ and $\tDel$ elements.  Following this, we relate the matter and vacuum values of the $X$s.
From Eq.~(\ref{lagrange2}), we have for the case $\alpha\neq\beta$, 
\begin{eqnarray}
\tDel_{ij}\tDel_{ik}\,\tX^i_{\alpha\beta} 
&=&[\tH^2-(\tlam_j+\tlam_k)\tH]_{\alpha\beta}\cr
&=&\tH_{\alpha\gamma}\tH_{\gamma\beta}
-(\tH_{\gamma\gamma}-\tlam_i)\tH_{\alpha\beta} 
\qquad (\alpha\neq\beta\neq\gamma,\,i\neq j\neq k) 
\label{xi2}
\end{eqnarray}
where we have used the fact that 
$Tr(\tH)\,(\equiv\tT)=\sum_{\alpha}\tH_{\alpha\alpha}=\sum_{i}\tlam_{i}$. 
Eq.~(\ref{xi2}) was obtained in \cite{kimura}, in a less straightforward manner. We note that on the right-hand side, all factors are matter-independent except for $(\tH_{\gamma\gamma}-\tlam_i)$. This factor is also
problematic, as it contains the $\tlam_i$, which are
not directly observable in neutrino oscillations. 
We deal with the both problems by the substitution in terms
of observable quantities:
\beq
(\tH_{\gamma\gamma}-\tlam_i)
=(H_{\gamma\gamma}-\frac{1}{3}{\cal T})-\ts_i+a\Dg
\label{dif2}
\eeq
where the 
\begin{equation}
\ts_i\equiv\frac{1}{3}(\tDel_{ij}+\tDel_{ik}) \quad (j \neq k\neq i),
\label{sdef}
\end{equation}
are the eigenvalues of the reduced matter-dependent Hamiltonian 
$(\tH-\frac{1}{3}\tT)$, and $\Dg$ is the ``$\gamma$'' element of the vector:
\begin{equation}
\underline{v}=(\frac{2}{3},-\frac{1}{3},-\frac{1}{3})
\label{dgdef}
\end{equation}
which breaks the symmetry between the three lepton flavours for $a\neq 0$. The matter-dependence in Eq.~(\ref{dif2}) has been isolated to its last two terms. Substitution from Eq.~(\ref{dif2}) in Eq.~(\ref{xi2}) shows that the quantities
\begin{eqnarray}
\tR&\equiv&\tDel_{ij}\tDel_{ik}\,\tX^i_{\alpha\beta}
+(a\Dg-\ts_i)\tH_{\alpha\beta}, 
\quad(\alpha\neq\beta\neq\gamma,\,i\neq j\neq k)\cr
&=&\tH_{\alpha\gamma}\tH_{\gamma\beta}
-(H_{\gamma\gamma}-\frac{1}{3}{\cal T})\tH_{\alpha\beta}\cr
&=&H_{\alpha\gamma}H_{\gamma\beta}
-(H_{\gamma\gamma}-\frac{1}{3}{\cal T})H_{\alpha\beta}
\label{newinv}
\end{eqnarray}
are matter-invariant, ie.~$\tR=R_{\alpha\beta}$ for all $\alpha\neq\beta$.

We can now relate the matter values $\tX^i$ and the vacuum values $X^i$:
\beq
\tX^i_{\alpha\beta} =
\frac
{\Delta_{ij}\Delta_{ik} X^i_{\alpha\beta}+
(\ts_i-\s_i-a\Dg)H_{\alpha\beta}}
{\tDel_{ij}\tDel_{ik}}
\qquad (\alpha\neq\beta\neq\gamma,\,j\neq k\neq i).
\label{xi4}
\eeq
The matter-invariants appearing in Eqs.~(\ref{newinv}) and (\ref{xi4}) may 
themselves be expanded in terms of vacuum values of $\Delta$s and $X$s:
\begin{eqnarray}
\tH_{\alpha\beta}=H_{\alpha\beta}
&=&\frac{1}{3}\sum^{cyclic}_{i,j,k}(\Delta_{ij}+\Delta_{ik})X^i_{\alpha\beta}
=\sum^{k~fixed}_{i}\Delta_{ik}X^i_{\alpha\beta}
\quad(\alpha\neq\beta)\label{HinX}\\
\tR=R_{\alpha\beta}&=&\frac{1}{3}\sum^{cyclic}_{i,j,k}\Delta_{ij}\Delta_{ik}X^i_{\alpha\beta}
=\frac{1}{3}\sum^{k~fixed}_{i\neq j\neq k}\Delta_{ik}(\Delta_{ij}+\Delta_{kj})X^i_{\alpha\beta}
\quad(\alpha\neq\beta)
\label{RinX}
\end{eqnarray}
where the former is obtained by summing 
Eq.~(\ref{set2}) over $k$ (with $a=0$) and using unitarity, 
and the latter by summing the first line of 
Eq.~(\ref{newinv}) over $i$ (again with $a=0$) and using the fact that 
$\sum_i \s_i=0$.
Using Eqs.~(\ref{newinv}) or (\ref{xi4}) with Eqs.~(\ref{HinX}) and
(\ref{RinX}) allows
the matter-modified values $\tX^i_{\alpha\beta}=\tU_{\alpha i}\tU^*_{\beta i}$ 
to be calculated without solving for the $\tU_{\alpha i}$ themselves
and without any dependence on the offset of the eigenvalues.

We remark that the $\tX^i_{\alpha\beta}$, $X^i_{\alpha\beta}$,
$H_{\alpha\beta}$ and $R_{\alpha\beta}$ ($\alpha\neq\beta$) are not 
observable, all having a common phase-convention dependence. 
Rather, the $\tX^i_{\alpha\beta}$
are to be considered as building blocks of the observables, $\tK$. The 
diagonal components, $\tX^i_{\alpha\alpha}=|\tU_{\alpha i}|^2$, are however 
observable; their calculation by a similar method is discussed in Appendix B.

It is now easy to calculate the
$T$-conserving and $T$-violating oscillation coefficients 
$\tK={\rm Re}(\tX^i_{\alpha\beta}\tX^{j*}_{\alpha\beta})$ 
and $\tilde{J}={\rm Im}(\tX^i_{\alpha\beta}\tX^{j*}_{\alpha\beta})$
in similar terms and exhibit their matter-dependences.
From Eq.\ (\ref{newinv}) we find:
\beq
\tK=\frac{\tilde{A}^k_{\gamma}\,|H_{\alpha\beta}|^2
    +\tilde{B}^k_{\gamma}\,{\rm Re}(H_{\alpha\beta}R^*_{\alpha\beta})
    +|R_{\alpha\beta}|^2}
     {\tDel_{12}\tDel_{23}\tDel_{31}\tDel_{ij}}
   \quad (\alpha\neq\beta),
\label{Kij}
\eeq
where all terms and factors are independently observable and the 
matter-dependence is confined to the coefficients
\begin{equation}
\tilde{A}^k_{\gamma}=\ts_i\ts_j+a\Dg\ts_k+a^2(\Dg)^2=\frac{1}{9}(-2\tDel^2_{ij}
+\tDel_{ki}\tDel_{kj})+\frac{1}{3}a\Dg(\tDel_{ki}+\tDel_{kj})+\frac{1}{3}a^2(\Dg+\frac{2}{3}),
\label{sisj}
\end{equation}
\begin{equation}
\tilde{B}^k_{\gamma}=-(\ts_k+2a\Dg)=-\frac{1}{3}(\tDel_{ki}+\tDel_{kj})-2a\Dg
\quad(i\neq j\neq k)
\label{siplussj}
\end{equation}
and the denominator. There is thus no dependence on the matter-modified mixing matrix elements, the matter-dependence entering only via the explicit $a$-dependent terms and the $\tDel_{ij}$, which are given by standard expressions in terms of vacuum parameters and the matter density in \cite{bargeretal}\cite{zaglauer}. Eq.~(\ref{Kij}) is similar to the exact formula for the $T$-even oscillations, Eq.~(\ref{kimura}), except that here, {\em Eq.~(\ref{Kij}) is composed of explicitly observable quantities}, and the matter-dependence has been isolated {\em for all} $\alpha\neq\beta$.

Similarly
\begin{eqnarray}
\tilde{J}&=&\frac{(\ts_i-\ts_j){\rm Im}(H_{\alpha\beta}R^*_{\alpha\beta})}
   {\tDel_{12}\tDel_{23}\tDel_{31}\tDel_{ij}}
   \qquad (\alpha\neq\beta)\cr
   &=&\pm\frac{{\rm Im}(\HHH)}{\tDel_{12}\tDel_{23}\tDel_{31}}
\label{J}
\end{eqnarray}
where we have used the relations
\begin{equation}
H_{\alpha\beta}R^*_{\alpha\beta}=(\HHH)^{(*)}-(H_{\gamma\gamma}-\frac{1}{3}{\cal T})|H_{\alpha\beta}|^2
\label{ReHR}
\end{equation}
and
\begin{equation}
(\ts_i-\ts_j)
=\tDel_{ij}.
\label{tsiminustsj}
\end{equation}

Eq.\ (\ref{J}) is simply the well-known result which leads to the NHS relation \cite{matterInv1}\cite{kimura}\cite{xing} of Eq.~(\ref{ImHHH}) above.

\vspace{7mm}
\ni {\bf 4 Exact Oscillation Probabilities in Terms of Vacuum Parameters}
\vspace{2mm}
\nl 
Using Eq.~(\ref{Kij}), we can now solve for the 
matter-dependence of the $\tK$ in terms of the vacuum $\Kijab$:
\begin{eqnarray}
\tDel_{12}\tDel_{23}\tDel_{31}\tDel_{ij}\tK
   &=&\Delta_{12}\Delta_{23}\Delta_{31}\Delta_{ij}K^{ij}_{\alpha\beta}
   +\kappa^{ij}_{\alpha\beta}
   \qquad (\alpha\neq\beta),
\label{KijCov1}
\end{eqnarray}
where
\begin{equation}
\kappa^{ij}_{\alpha\beta}
   =(\tilde{A}^k_{\gamma}-A^k_{\gamma})|H_{\alpha\beta}|^2
     +(\tilde{B}^k_{\gamma}-B^k_{\gamma}){\rm Re}(H_{\alpha\beta}R^*_{\alpha\beta})
   \quad (\alpha\neq\beta\neq\gamma,~i\neq j\neq k)
\label{kappa}
\end{equation}
with $\tilde{A}^k_{\gamma}$ and $\tilde{B}^k_{\gamma}$ given in 
Eqs.~(\ref{sisj}) and (\ref{siplussj}) and
$A^k_{\gamma}=\frac{1}{9}(-2\Delta^2_{ij}+\Delta_{ik}\Delta_{jk})$
and $B^k_{\gamma}=-\frac{1}{3}(\Delta_{ki}+\Delta_{kj})$
(which do not depend on $\gamma$).
Equation (\ref{KijCov1}) is the exact $T$-even analogue of the 
$T$-odd invariance, Eq.~(\ref{ImHHH}). It is slightly more 
complicated in the sense that the matter-modified values $\tK$ differ 
from the $\Kijab$ by an inhomogeneous term 
$\kappa^{ij}_{\alpha\beta}/\tdiscr\tDel_{ij}$, 
as well as by a scale factor. The $\kappa^{ij}_{\alpha\beta}$ clearly vanish 
in the limit $a\rightarrow 0$, as they should.

In order to complete our specification of the $\tK$ 
in terms of the vacuum $\Kijab$, it is necessary to
give $|H_{\alpha\beta}|^2$ and ${\rm Re}(H_{\alpha\beta}R^*_{\alpha\beta})$
in Eq.~(\ref{kappa}) in these terms. The former is given by \cite{matterInv2}:
\begin{equation}
-\sum^{cyclic}_{(ij)}\tDel^2_{ij}\tK=|\tH_{\alpha\beta}|^2
=|H_{\alpha\beta}|^2=-\sum^{cyclic}_{(ij)} 
\Delta^2_{ij}\Kijab, 
\quad (\alpha \neq \beta) \,,
\label{Hab}
\end{equation}
while the latter may be derived from Eqs.~(\ref{HinX}) and (\ref{RinX}) as 
discussed in Appendix C:
\begin{equation}
{\rm Re}(H_{\alpha\beta}R^*_{\alpha\beta})
=-\frac{1}{3}\sum^{cyclic}_{i,j,k}\Delta^2_{ij}(\Delta_{ik}+\Delta_{jk})K^{ij}_{\alpha\beta}
\qquad(\alpha\neq\beta).
\label{ReHR2}
\end{equation}
Eqs.~(\ref{KijCov1})-(\ref{ReHR2}) are the main result of this paper, which, along with Eqs.~(\ref{sisj}) and (\ref{siplussj}), allow the $\tK$ to be calculated in terms of the vacuum parameters $\Delta_{ij}$ and $K^{ij}_{\alpha\beta}$, the $\tDel_{ij}$ and the matter parameter, $a$, thereby avoiding the need to find the matter-modified MNS mixing matrix. These formulae are, furthermore, all convention-independent.

\begin{figure}[!ht]
\begin{center}
\mbox{\epsfig{file=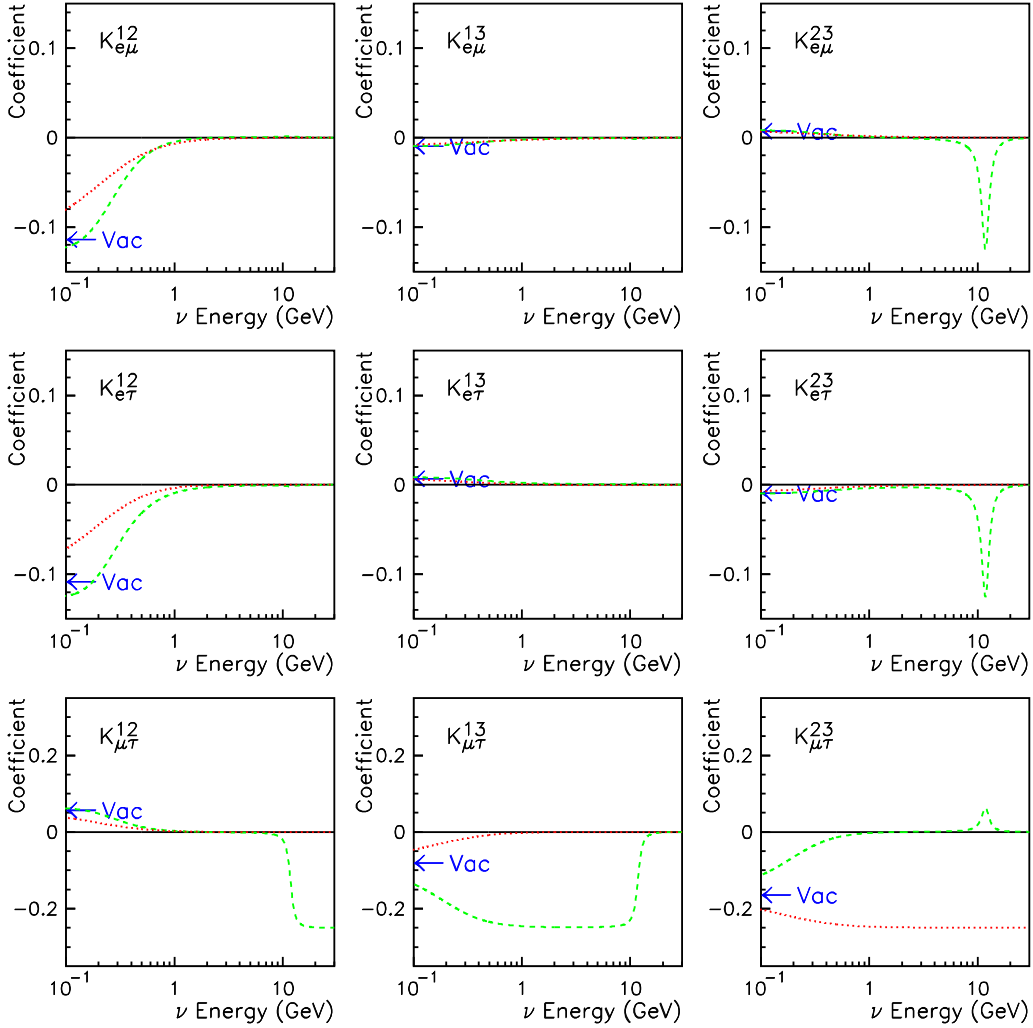,width=14cm}}
\end{center}
\caption{The nine coefficients $\tK$, $\alpha\neq\beta$, $i\neq j$, for neutrinos (dashed lines) and antineutrinos (dotted lines) traversing the Earth's mantle, as functions of the neutrino energy. We take 
$\Delta m^2_{12}=5.0\times 10^{-5}$ eV$^2$, $\Delta m^2_{13}=2.5\times 10^{-3}$ eV$^2$, $\sin{\theta_{12}}=0.58$, $\sin{\theta_{23}}=0.71$, 
$\sin{\theta_{13}}=0.05$ and $\delta=\pi/4$. The corresponding vacuum values are indicated by an arrow.}
\label{fig1}
\end{figure}

In Fig.~(\ref{fig1}) 
we plot the nine $\tK$, $\alpha\neq\beta$, $i\neq j$, for neutrinos and antineutrinos traversing the Earth's mantle, as functions of the neutrino energy, calculated using Eqs.~(\ref{KijCov1})-(\ref{ReHR2}).
We take $\Delta m^2_{12}=5.0\times 10^{-5}$ eV$^2$, 
$\Delta m^2_{13}=2.5\times 10^{-3}$ eV$^2$, 
$\sin{\theta_{12}}=0.58$, $\sin{\theta_{23}}=0.71$, 
$\sin{\theta_{13}}=0.05$ and $\delta=\pi/4$.
The similarity of the first and second rows reflects the approximate 
$\numu$-$\nutau$ symmetry \cite{mutau} of the vacuum MNS matrix.
Except near a matter-resonance, the smallness of the values in the 
upper-right 2x2 sub-block reflects the smallness of $|U_{e3}|$.

From Eq.~(\ref{prob}), we can write the exact expression for the appearance probabilities in uniform density matter:
\begin{eqnarray}
\tP(\nu_{\alpha} \rightarrow \nu_{\beta})=|\tcA_{\alpha\beta}|^2 &=& 
-4\sum_{i<j}\frac{\Delta_{12}\Delta_{23}\Delta_{31}\Delta_{ij}K^{ij}_{\alpha\beta}
   +\kappa^{ij}_{\alpha\beta}}{\tdiscr\tDel_{ij}}\sin^2{(\tDel_{ij}L/2)} \cr
&+&8\frac{\discr}{\tdiscr}J\sin{(\tDel_{12}L/2)}
\sin{(\tDel_{23}L/2)}\sin{(\tDel_{31}L/2)}\cr 
&&\label{probMat}
\end{eqnarray}
($\alpha\neq\beta$), where $\kappa_{ij}$ is given by Eq.~(\ref{kappa}) in
terms of vacuum quantities and the $\tilde{A}^k_{\gamma}$ 
and $\tilde{B}^k_{\gamma}$, 
which in turn depend only on $a$ and the $\tDel_{ij}$.
Although we have calculated only appearance probabilities in 
Eq.~(\ref{probMat}), survival probabilities are calculable in similar terms
directly from them using unitarity.
This completes our derivation of the exact matter-covariant 
formulation of neutrino oscillation probabilities in uniform density
matter, in terms of vacuum oscillation parameters and the matter density.

Our formulae also hold for antineutrino oscillations. For antineutrinos, the signs of $J$ and of $a$ are opposite to those for neutrinos. These sign changes alter the effective Hamiltonian, and the values of the eigenvalue-differences, $\tDel_{ij}$, are changed in our formulae. In Fig.~\ref{fig1} the consequent differences between the antineutrino and neutrino oscillation coefficients are clearly seen. We discuss this more in Appendix A. 

\newpage
\vspace{7mm}
\ni {\bf Appendix A: Matter-Effects for Antineutrino vs.\ Neutrino}
\vspace{2mm}
\nl
Since matter is inherently CP-violating, it affects antineutrinos 
differently than neutrinos.  This can be summarised by noting that 
the sign of $a$ is negative for antineutrinos, positive for neutrinos.
Thus, any matter invariant is also a neutrino-antineutrino invariant,
while any matter dependence breaks the neutrino-antineutrino symmetry.
The explicit breaking of this symmetry is readily obtained
from some of our derived formulae. From Eq.\ (\ref{xi4}) we get
\bea
\left[{\tDel_{ij}\tDel_{ik}}\tX^i_{\alpha\beta}\right]_\nu -
\left[{\tDel_{ij}\tDel_{ik}}\tX^i_{\alpha\beta}\right]_\nubar 
&=&([\ts_i]_\nu-a\Dg-[\ts_i]_\nubar-a\Dg)H_{\alpha\beta}
\quad (\alpha\neq\beta,\,j\neq k\neq i)\label{Xabnnbar}\cr
&=&\left\{\frac{1}{3}([\tDel_{ij}+\tDel_{ik}]_\nu -
[\tDel_{ij}+\tDel_{ik}]_\nubar)
-2a\Dg\right\}H_{\alpha\beta}\,.\cr
&~&
\eea
From Eq.\ (\ref{J}) we get 
\beq
\left[
\tDel_{12}\tDel_{23}\tDel_{31}\tilde{J}
\right]_\nu =
-\left[
\tDel_{12}\tDel_{23}\tDel_{31}\tilde{J}
\right]_\nubar.
\label{Jnnbar}
\eeq
Similar but longer expressions may be written down for
\beq
\left[\tDel_{12}\tDel_{23}\tDel_{31}\tDel_{ij}\tK\right]_\nu -
\left[\tDel_{12}\tDel_{23}\tDel_{31}\tDel_{ij}\tK\right]_\nubar
=\left[\kappa^{ij}_{\alpha\beta}\right]_{\nu}
-\left[\kappa^{ij}_{\alpha\beta}\right]_{\nubar}
\eeq
using Eqs.\ (\ref{KijCov1}) and (\ref{kappa}) (the terms quadratic in $a$
cancel) and for
\beq
\left[
{\tDel_{ij}\tDel_{ik}} |\tU_{\alpha i}|^2
\right]_\nu -
\left[
{\tDel_{ij}\tDel_{ik}} |\tU_{\alpha i}|^2
\right]_\nubar
\eeq
using Eq.\ (\ref{Umatvac}).

\vspace{7mm}
\ni \boldmath {\bf Appendix B: Matter-Vacuum Relation for $|\tU_{\alpha i}|^2$}
\unboldmath
\vspace{2mm}
\nl
The starting point is the expansion of the 
diagonal element of Eq.\ (\ref{lagrange2}):
\bea
{\tDel_{ij}\tDel_{ik}}\tX^i_{\alpha\alpha} 
&=& \sum_{\sigma=e,\mu,\tau}
	(\tH-\tlam_j)_{\alpha\sigma}(\tH-\tlam_k)_{\sigma\alpha}
 	\quad (j \neq k \neq i) \\
&=& (\tH_{\alpha\alpha}-\tlam_j)(\tH_{\alpha\alpha}-\tlam_k)
+|H_{\alpha\beta}|^2+|H_{\alpha\gamma}|^2
	\quad (\alpha \neq \beta \neq \gamma)\,.
\label{lagrange3}
\eea
Inputting Eq.\ (\ref{dif2}) for $(\tH_{\alpha\alpha}-\tlam_j)$
then isolates the matter dependence.
The result is
\beq
{\tDel_{ij}\tDel_{ik}} |\tU_{\alpha i}|^2 =
{\Delta_{ij}\Delta_{ik}} |U_{\alpha i}|^2+(\tA^i_{\alpha}-A^i_{\alpha})
-(\tB^i_{\alpha}-B^i_{\alpha})(H_{\alpha\alpha}-\frac{1}{3}{\cal T})
\label{Umatvac}
\eeq
where the coefficients are the same as those in $\kappa^{ij}_{\alpha\beta}$,
Eq.~(\ref{kappa}).
The factor $(H_{\alpha\alpha}-\frac{1}{3}{\cal T})$ can also be put in 
terms of our vacuum observables using the appropriate diagonal component of the
matrix equation:
\beq
(H-\frac{1}{3}{\cal T})=\frac{1}{3}\sum^{cyclic}_{i,j,k}(\Delta_{ij}+\Delta_{ik})X^i
\label{reduced}
\eeq
which is obtained by summing Eq.~(\ref{set2}) over $k$ (taking the vacuum
case).

\vspace{7mm}
\ni \boldmath{\bf Appendix C: Matter Invariants in Terms of
$K^{ij}_{\alpha\beta}$ and $|U_{\alpha i}|^2$}\unboldmath
\vspace{2mm}
\nl The derivation of Eq.~(\ref{ReHR2}) follows straightforwardly 
from Eqs.~(\ref{HinX}) and (\ref{RinX}), utilising the useful relations 
between the $K^{ij}_{\alpha\beta}$ and the $|X^i_{\alpha\beta}|^2$:
\beq
|X^i_{\alpha\beta}|^2=-K^{ij}_{\alpha\beta}-K^{ik}_{\alpha\beta} 
\quad \forall~\alpha\neq\beta, \quad(i\neq j\neq k)
\label{Xisq}
\eeq
and
\beq
K^{ij}_{\alpha\beta}=\frac{1}{2}(|X^k_{\alpha\beta}|^2
-|X^i_{\alpha\beta}|^2-|X^j_{\alpha\beta}|^2)
\quad \forall~\alpha\neq\beta, \quad(i\neq j\neq k)
\label{KijinXisq}
\eeq
which are themselves easily derived from the unitary condition 
$\sum_iX^i_{\alpha\beta}=0$.

These can also be used to find the matter-invariant
\beq
|R_{\alpha\beta}|^2=-\frac{1}{9}\sum^{cyclic}_{i,j,k}
\Delta^2_{ij}(\Delta_{ik}+\Delta_{jk})^2K^{ij}_{\alpha\beta}
\label{RabsqinK}
\eeq
which, in addition to Eqs.~(\ref{Hab}) and (\ref{ReHR2}), completes the 
set of matter-invariants used in Eq.~(\ref{Kij}) (we did not need to specify
these in the main text, because we used the substitution of 
$K^{ij}_{\alpha\beta}$ instead to find Eqs.~(\ref{KijCov1}) and (\ref{kappa})).

We can also use Eqs.~(\ref{Xisq}) and (\ref{KijinXisq}) to find 
the set of three matter invariants used in Eq.~(\ref{Kij}), instead 
in terms of $|X^i_{\alpha\beta}|^2=|U_{\alpha i}|^2|U_{\beta i}|^2$:
\beq
|H_{\alpha\beta}|^2=\sum^{cyclic}_{i,j,k}
\Delta_{ij}\Delta_{ik}|X^i_{\alpha\beta}|^2
\label{HabsqinXsq}
\eeq
\beq
{\rm Re}(H_{\alpha\beta}R^*_{\alpha\beta})=\frac{1}{6}\sum^{cyclic}_{i,j,k}
\Delta_{ij}\Delta_{ik}(\Delta_{ik}+\Delta_{jk})|X^i_{\alpha\beta}|^2
\label{ReHRinXsq}
\eeq
\beq
|R_{\alpha\beta}|^2=\frac{2}{9}\sum^{cyclic}_{i,j,k}
\Delta_{ij}\Delta_{ik}(\Delta_{ij}\Delta_{ik}-2\Delta^2_{jk})|X^i_{\alpha\beta}|^2.
\label{RabsqinXsq}
\eeq
Substituting these into Eq.~(\ref{Kij}) yields a formula for the 
$\tK$ which depends only on the moduli-squared, $|U_{\alpha i}|^2$, of
elements of the vacuum MNS matrix (in addition to the $\tDel_{ij}$ and the 
matter-density, $a$). This alternative formulation may be considered more
convenient to use than the one using the vacuum $\Kijab$.

\vspace{7mm}
\ni {\bf Acknowledgements}
\vspace{2mm}
\nl This work was supported by the UK Particle Physics and Astronomy 
Research Council (PPARC) and by the U.S. Department of Energy under 
grant number DE-FG05-85ER40226. T.W. acknowledges support from the 
CCLRC Rutherford Appleton Laboratory.

\end{document}